\documentclass{article}

\usepackage[preprint, nonatbib]{neurips_2021}

\usepackage[utf8]{inputenc} 
\usepackage[T1]{fontenc}    
\usepackage[pagebackref=true,breaklinks=true,colorlinks,bookmarks=false]{hyperref}  
\usepackage{url}            
\usepackage{booktabs}       
\usepackage{nicefrac}       
\usepackage{microtype}      
\usepackage{xcolor}         
\usepackage{wrapfig}

\usepackage{cite}
\usepackage[numbers]{natbib}
\usepackage{amsmath,amssymb,amsfonts}
\usepackage[noend]{algorithmic}
\usepackage{graphicx}
\usepackage{textcomp}
\usepackage{algorithm}

\newtheorem{remark}{Remark}

\newtheorem{definition}{Definition}
\newtheorem{problem}{Problem}

\newcommand{\FS}[2]{\displaystyle\frac{#1}{#2}}
\newcommand\at[2]{\left.#1\right|_{#2}}

\newenvironment{myitemize}{\begin{list}{$\bullet$}
{\setlength{\topsep}{1mm}
\setlength{\itemsep}{0.25mm}
\setlength{\parsep}{0.25mm}
\setlength{\itemindent}{0mm}
\setlength{\partopsep}{0mm}
\setlength{\labelwidth}{15mm}
\setlength{\leftmargin}{4mm}}}{\end{list}}

\title{Verification in the Loop: Correct-by-Construction Control  Learning with Reach-avoid Guarantees}

\author{%
  Yixuan Wang \\
  Northwestern University\\
  \texttt{wangyixu14@gmail.com} \\
   \And
   Chao Huang \\
   Liverpool University \\
   \texttt{thomashuangchao@gmail.com} \\
   \AND
   Zhaoran Wang \\
   Northwestern University \\
   \texttt{zhaoranwang@gmail.com } \\
   \And
   Zhilu Wang \\
   Northwestern University \\
   \texttt{zhiluwang2018@u.northwestern.edu} \\
   \And
   Qi Zhu \\
   Northwestern University \\
   \texttt{qzhu@northwestern.edu } \\
}

\begin{document}

\maketitle
 
\begin{abstract}
In the current control design of safety-critical autonomous systems, formal verification techniques are typically applied \emph{after} the controller is designed to evaluate whether the required properties (e.g., safety) are satisfied. However, due to the increasing system complexity and the fundamental hardness of designing a controller with formal guarantees, such an open-loop process of \emph{design-then-verify} often results in many iterations and fails to provide the necessary guarantees. 
In this paper, we propose a correct-by-construction control learning framework that integrates the verification into the control design process in a closed-loop manner, i.e., \emph{design-while-verify}. 
Specifically, we leverage the verification results (computed reachable set of the system state) to construct feedback metrics for control learning, which measure how likely the current design of control parameters can meet the required reach-avoid property for safety and goal-reaching. We formulate an optimization problem based on such metrics for tuning the controller parameters, and develop an approximated gradient descent algorithm with a difference method to solve the optimization problem and learn the controller. 
The learned controller is formally guaranteed to meet the required reach-avoid property. By treating verifiability as a first-class objective and effectively leveraging the verification results during the control learning process, our approach can significantly improve the chance of finding a control design with formal property guarantees. This is demonstrated via a set of experiments on both linear and non-linear systems that use model-based or neural network based controllers. 
\end{abstract}

\section{Introduction}


Safety-critical autonomous systems, such as avionics systems~\cite{cofer2020run} and self-driving vehicles~\cite{koopman2017autonomous}, often operate in highly dynamic environments with significant uncertainties and disturbances. It is critical yet challenging to formally ensure their safety, especially for the control and decision making modules.   
Thus, while there has been increasing interest in applying machine learning techniques (especially neural network based ones such as deep reinforcement learning~\cite{sutton2018reinforcement}) to control and general decision making, their adoption in safety-critical systems is hindered by the challenges in formally ensuring system properties~\cite{wang2021cocktail,ivanov2020case}.  

In this work, we address system safety and goal-reaching ability in control design with a \emph{reach-avoid property}~\cite{margellos2011hamilton, jin2020neural}, which intuitively represents whether the system can ``reach its goal without entering unsafe states'' (formal definition in Section~\ref{sec:problem_formulation}).
It is a fundamentally hard problem to design a controller with formal guarantees for such property. Even in linear systems, the similar ``hyper-plane hitting problem'' is proved to be NP-hard and it is unclear whether the problem is decidable or not~\cite{blondel2000survey}. The complexity continues to increase for non-linear and hybrid systems~\cite{henzinger1998s}. Moreover, for emerging neural network based controllers, the typical black-box view of their behavior makes safety and stability verification extremely difficult~\cite{xiang2018reachability,aydinoglu2020stability}, let alone synthesize them with formal guarantees. 

The current process for controller design and verification follows an open-loop \emph{design-then-verify} pattern. The experts first design a controller using either model-based methods such as model predictive control~\cite{agachi20162} and linear quadratic regulator~\cite{bemporad2002explicit}, or model-free approaches such as reinforcement learning (RL) with neural networks~\cite{lillicrap2016continuous, schulman2017proximal}. They then leverage formal verification tools~\cite{ivanov2019verisig, huang2019reachnn, chen2013flow} to evaluate whether the designed controller satisfies the required properties. However, due to the above-mentioned difficulty of designing a controller with formal guarantees, such a process often results in many iterations between design and verification, with numerous attempts in tuning the control design and parameter settings, and may still fail to provide the necessary guarantees. For 
neural network based controllers, this could be even more challenging, as tuning the design and learning parameters often has an unpredictable impact on the properties of the learned  controller~\cite{henderson2018deep}. 



In this work, to address the above challenges, we propose a correct-by-construction control learning framework that integrates verification in a closed-loop manner, i.e., \emph{design-while-verify}, to formally guarantees that the learned controller satisfies the required reach-avoid property.
In our framework, we leverage the verification results, particularly the computed reachable set of the system state, to construct two different types of feedback metrics that reflect the system's potential ability to meet the reach-avoid property (one based on geometric distance, and one on Wasserstein distance). 
We then formulate the control learning as an optimization problem of the control parameters based on either metric, and develop an approximated gradient descent algorithm with a difference method for tuning the control parameters until a feasible solution is obtained or iteration limit is reached.
Our approach can be applied to both model-based controllers and neural network based ones, and formally guarantees that the learned controller can meet the required reach-avoid property.

\noindent
\textbf{Related work:} Our work is related to the safety verification of controlled dynamical systems~\cite{ivanov2019verisig, huang2019reachnn, chen2013flow, dutta2019reachability}, which typically relies on the computation of the reachable set containing all possible states that the system may visit within a time horizon. 
Our approach leverages these verification tools (called \emph{verifiers} in the rest of the paper), and develops novel metrics and method to integrate them into the control design process. 
Falsification is another technique that can be leveraged for closed-loop controller design~\cite{dreossi2019compositional,fremont2020formal,dreossi2019verifai}. Specifically, the falsifier searches for counter-examples to the required property and proposes measures to remedy failed cases. 
However, the falsification-driven process does not provide formal guarantees as in our approach. 

Our framework is related to control learning with certificates, such as barrier certificate with safety consideration and Lyapunov function (certificate) with stability concerns~\cite{qin2021learning,jin2020neural, zhao2020synthesizing, chang2020neural}. However, such data-driven approaches for synthesizing certificates need to sample system states from the unsafe region, which is not practical in most safety-critical systems. Moreover, the synthesized certificates and the learned controllers still need to be verified to provide any formal guarantees. In another word, they also follow the traditional \emph{design-then-verify} paradigm.
Our approach is also related to safe learning for dynamical systems~\cite{garcia2015comprehensive}, such as safe reinforcement learning with formal methods~\cite{fulton2018safe} and shielding~\cite{alshiekh2018safe, li2020robust}, where the agent is restricted to explore within an action set that is known a priori to be safe for the current state. In contrast, our framework does not constrain the action space and ensures system safety by assessing the reachable set -- the system always stays within the safe region during learning and is guaranteed to be safe with the learned controller.

In summary, our work makes the following novel contributions:

\begin{myitemize}
\item We propose a correct-by-construction control learning framework that integrates verification in a closed-loop \emph{design-while-verify} manner, which formally ensures that with the learned controller, the system satisfies the reach-avoid property for safety and goal-reaching.
\item Our framework includes novel formulation of the verification-in-the-loop control learning problem based on two different metrics (using geometric or Wasserstein distance) and an approximate gradient descent algorithm with a difference method for solving the problem. 
\item Our approach can be applied to both linear and non-linear systems under traditional model-based or emerging neural network based controllers. Experiments on a linear adaptive cruise control system and non-linear Van der Pol's oscillator system demonstrate that our approach significantly outperforms the baseline methods in convergence rate, safe control rate, goal-reaching rate, and ability to provide formal guarantees.
\end{myitemize}

The paper is organized as follows. Section~\ref{sec:problem_formulation} presents the system model. Section~\ref{sec:methodology} introduces our verification-in-the-loop control learning approach, including the definition of feedback metrics on verification results, the optimization problem formulation for control learning, the approximate gradient descent algorithm for solving the problem, and its optimality analysis for the Wasserstein metric. 
Experiments, further discussion and conclusion are presented in Sections~\ref{sec:experiment}, \ref{sec:discussion} and \ref{sec:conclusion}, respectively. 

\section{System Model}\label{sec:problem_formulation}

\paragraph{System Dynamics and Controller:} We consider a continuous system that can be expressed as a tuple $(X, U, f,  \kappa_\theta, X_0, \delta)$. Specifically, the dynamics of the continuous system is modeled as
\begin{equation}\label{eq:sysmodel}
    \dot{x} = f(x,u),
\end{equation}
where $x\in X \in \mathbb{R}^n$ is the system state vector and $X$ is the system state space. $u \in U \in \mathbb{R}^{m}$ is the control input variable, where $U$ is the control input space. $f:X \times U \rightarrow X$ is a locally Lipschitz-continuous function that can be either linear or non-linear. $X_{0}$ is a set containing all possible initial states $x(0)$. 

Such a system can be controlled by a feedback controller $\kappa_{\theta}: X \rightarrow U$, which is parameterized by $\theta$ in the following way. Given a sampling period $\delta$, the controller $\kappa$ reads the system state $x(i\delta)$ at time $t=i\delta (i = 1, 2, \cdots)$, and computes the control input as $u(i\delta) = \kappa_\theta(x(i\delta))$. Then, the system state evolves as $\dot{x} = f(x, u(i\delta))$ within the time slot $[i\delta, (i+1)\delta]$. 

\begin{remark}

Our approach can be applied to a variety of controller types, such as linear controllers, polynomial controllers, and fully-connected feed-forward neural network controllers that are expressed as
\begin{displaymath}
\kappa_\theta(x) = \kappa_L(\kappa_{L-1}\cdots(\kappa_1(x;W_1, b_1);W_{L-1}, b_{L-1});W_L, b_L),
\end{displaymath}
where $\theta=\{W_{i}, b_{i}(i=1,\cdots, L)\}$, $W_{i}, b_{i}$ are the weights and bias parameters for layer $i$ and $L$ is the number of layers. 
\end{remark}

\paragraph{Flow and Reach-avoid Property:} 
A \textit{flow} function $\varphi(x(0), t): X_0 \times \mathbb{R}_{+} \rightarrow X$ maps some initial state $x(0)$ to the system state $\varphi(x(0), t)$ at time $t$. Mathematically, $\varphi$ satisfies 1) $\varphi(x(0), 0) = x(0)$ 2) $\varphi$ is the solution of the $\dot{x} = f(x, u(i\delta))$ in the time interval $t \in [i\delta, i\delta+\delta]$ 3) $u(i\delta) = \kappa_{\theta}(\varphi(x(0), i\delta))$, $\forall i = 1, 2, \cdots$
 Based on the \textit{flow} definition, the system reach-avoid property is defined as follows.



\begin{definition}
(\textbf{Reach-avoid property}) Starting from an initial state $x(0)$, the system is considered as meeting the reach-avoid property if and only if its \textit{flow} 1) never enters into an unsafe set $X_{u}$(safety) and 2) reaches a goal set $X_g$(goal-reaching) within a finite time horizon $T$. 
\begin{displaymath}
\forall \ T \geq  t\geq 0, \ \varphi (x(0), t) \cap X_{u} = \emptyset(safety), \exists \ 0 \leq t' \leq T, \ \varphi(x(0), t') \cap X_g \neq \emptyset(goal-reaching)
\end{displaymath}
\end{definition}


\paragraph{Verifier and Control Learning:} We consider a verifier as a formal tool $\Psi(f, X_0, \kappa_{\theta})$ that takes input of system dynamics $f$, initial state set $X_0$, and controller $\kappa_{\theta}$, and outputs the feedback concerning reach-avoid property. 
Leveraging such formal verifier, we define our closed-loop control learning problem with reach-avoid guarantee as the following.     

\begin{problem}
(\textbf{Verification-in-the-loop control learning}) Given a continuous control system described as \eqref{eq:sysmodel}, find a feasible solution of controller parameters $\theta$ and initial region $X_I \subseteq X_0$ with the feedback from verifier $\Psi(f, X_0, \kappa_\theta)$, such that the reach-avoid property is satisfied for every possible $x(0) \in X_I \subseteq X_0$ with controller $\kappa_\theta$. 
\end{problem}

\section{Verification-in-the-Loop Control Learning}\label{sec:methodology}
We leverage the feedback from the verifier to guide the control learning process. Our verification-in-the-loop approach includes the following major components: the computation of the system state reachable set from the verifier (Section~\ref{subsection:reachable_set}); the two different definitions of a distance metric over the reachable set for evaluating the current control design and the formulation of an optimization problem for control learning (Section~\ref{subsection:metrics}); and an approximated gradient descent algorithm for solving the optimization problem, including the computation of an initial state set for ensuring goal-reaching and the algorithm optimality analysis for Wasserstein metric (Section~\ref{subsection:learning_algorithm}).



\subsection{Verifier Reachable Set Computation}
\label{subsection:reachable_set}

During the verification-in-the-loop control learning process, the verifier $\Psi(f, X_0, \kappa_\theta)$ computes a reachable set of the system state based on the current controller design $\kappa_{\theta}$, defined as follows.
\begin{definition}
A state $x_r$ of system $(X, U, f,  \kappa_\theta, X_0, \delta)$ is called reachable at time $t \geq 0$, if and only if there $\exists \ x(0) \in X_0$ such that $x_r = \varphi(x(0), t)$ under controller $\kappa_{\theta}$. The reachable set $X_r^{T}$ with time horizon $T$ for initial set $X_0$ is defined as
\begin{displaymath}
X_{r}^{T} = \{ \varphi(x(0), t) \ | \ \forall \ x(0) \in X_{0},  \ \forall \  0 \leq t \leq T\}
\end{displaymath}
\end{definition}

For computing this reachable set, we consider two cases: linear systems under linear controllers, and non-linear systems under  non-linear controllers such as neural network based ones. 


\paragraph{Linear System with Linear Controller:} For a linear time-invariant (LTI) system as
\begin{displaymath}
 \dot{x} = Ax + Bu,
\end{displaymath}
its reachable set under a linear controller within a finite time interval can be evaluated recursively~\cite{wang_date_21}. Specifically, we consider the discretized LTI system as $x[t+1] = A_d x[t] + B_d u[t]$ with a linear feedback controller $u[t]=\theta^T x[t]$, where $A_d = e^{A\delta}$, $B_d = \int^{\delta}_{0} e^{At} B dt$ with sampling period $\delta$. Note that for continuous LTI systems, as long as the controller is periodically updated and zero-order hold is applied in each period, it can always be discretized. The initial set $X_0$ is considered as a polyhedron. In this case, the reachable set of each time step $t$, denoted as $X_r[t]$, is also a polyhedron, and can be derived recursively from $X_0$ by polyhedron operation $X_r[t+1] = (A_d + B_d \theta) X_r[t]$ with $X_r[0] = X_0$. The overall reachable set can be obtained as $X_r^{T} = \bigcup_{t=0}^{T} X_r[t]$. 


\paragraph{Non-linear System with Neural Network Controller:} For a non-linear system with a neural network controller, we leverage the method from ReachNN~\cite{huang2019reachnn}. First, we apply the overly function approximation for the neural network by Bernstein polynomials with bounded error.  
\begin{definition}
(Neural network with function approximator) Let $d=(d_{1}, \cdots, d_{n}) \in \mathbb{R}^{n}$ and $\kappa_{\theta}$ be a continuous neural network controller over variables $x \in [0, 1]^n$. The polynomial related to the controller $\kappa_{\theta}$
\begin{displaymath}
B_{\kappa_{\theta}, d}(x) = \sum_{\substack{0\leq a_{j} \leq d_{j} \\ j=\{1, 2,\cdots, n\}}} \kappa_{\theta} \left(\frac{a_{1}}{d_{1}},\cdots,\frac{a_{n}}{d_{n}} \right) \prod_{j=1}^{n} \left (\tbinom{d_{j}}{a_{j}}x_{j}^{a_{j}}(1-x_{j})^{d_{j}-a_{j}} \right)
\end{displaymath}
is called Bernstein polynomials approximator of $\kappa_{\theta}$ under degree $d$. The Bernstein polynomial approximator along with an error bound $\epsilon$ ensures that the output range of neural network under some reachable set $X_r$ is bounded as
 \begin{displaymath}
 u = \kappa_{\theta}(x) \in B_{\kappa_{\theta}, d}(x) + [-\epsilon, \epsilon], \forall x \in X_r
\end{displaymath}
\end{definition}

In this way, the neural network controller is transformed into polynomials while the approximation error is treated as the external disturbances to the system. This enables us to efficiently compute an over-approximation of the reachable set for neural network-controlled systems. 

\subsection{Distance Metric Definitions over Reachable Set and Control Learning Formulation}
\label{subsection:metrics}

We define two different types of metrics for evaluating the current control design based on the computed reachable set from the verifier, one based on the intuitive geometric distance and one on the Wasserstein distance for its convexity. 

\paragraph{Geometric Distance based Metrics:}

\begin{figure}[tbp]
\centering
\begin{minipage}[t]{0.48\textwidth}
\centering
\includegraphics[width=1.08\linewidth]{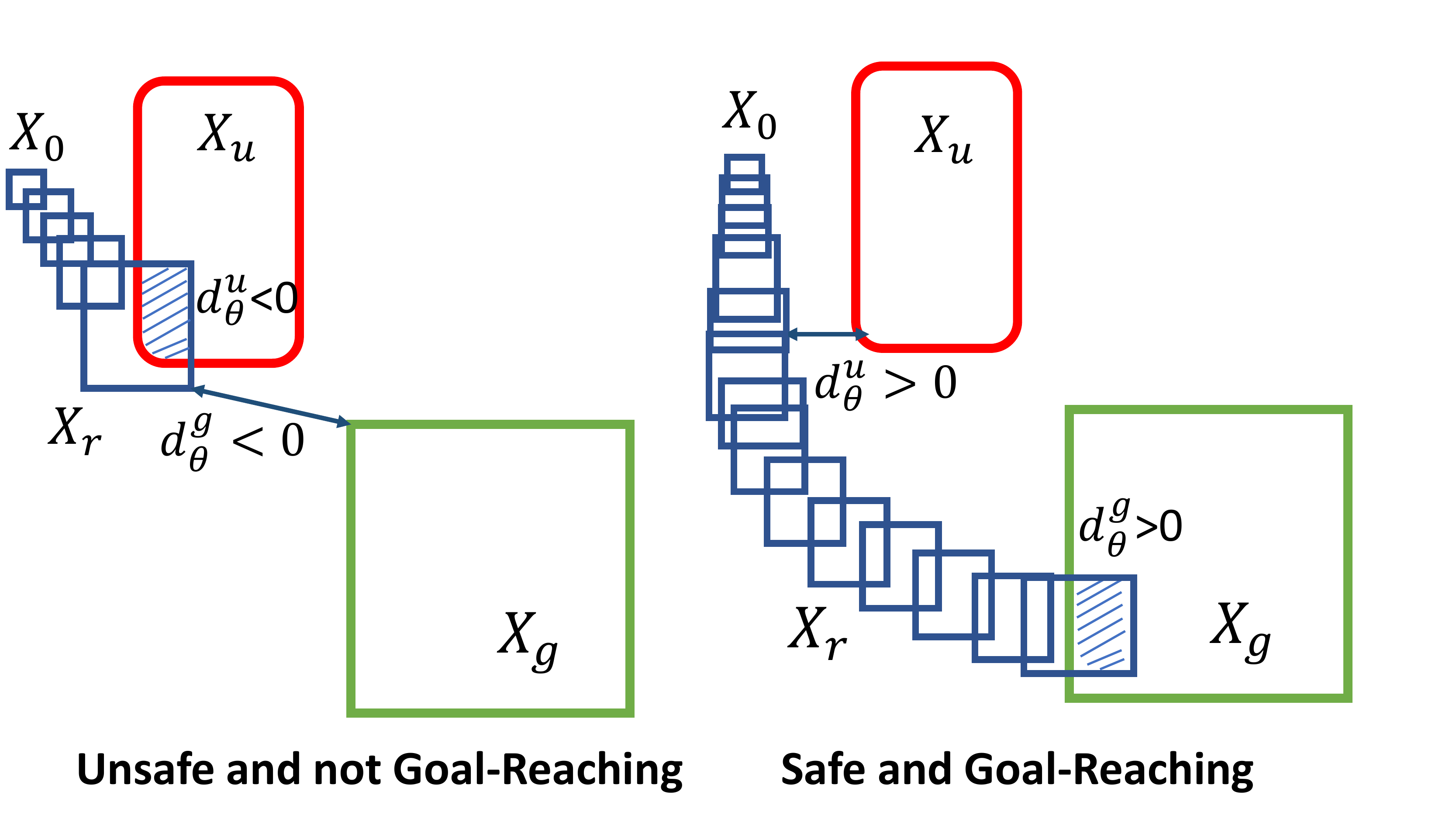}
\caption{Geometric distances $d_{\theta}^{u}$, $d_{\theta}^{g}$ for safe and goal-reaching properties defined on reachable set (blue) with respect to unsafe region (red) and goal set (green).}
\label{fig:metrics}
\end{minipage}
\quad
\begin{minipage}[t]{0.48\textwidth}
\centering
\includegraphics[width=1.11\linewidth]{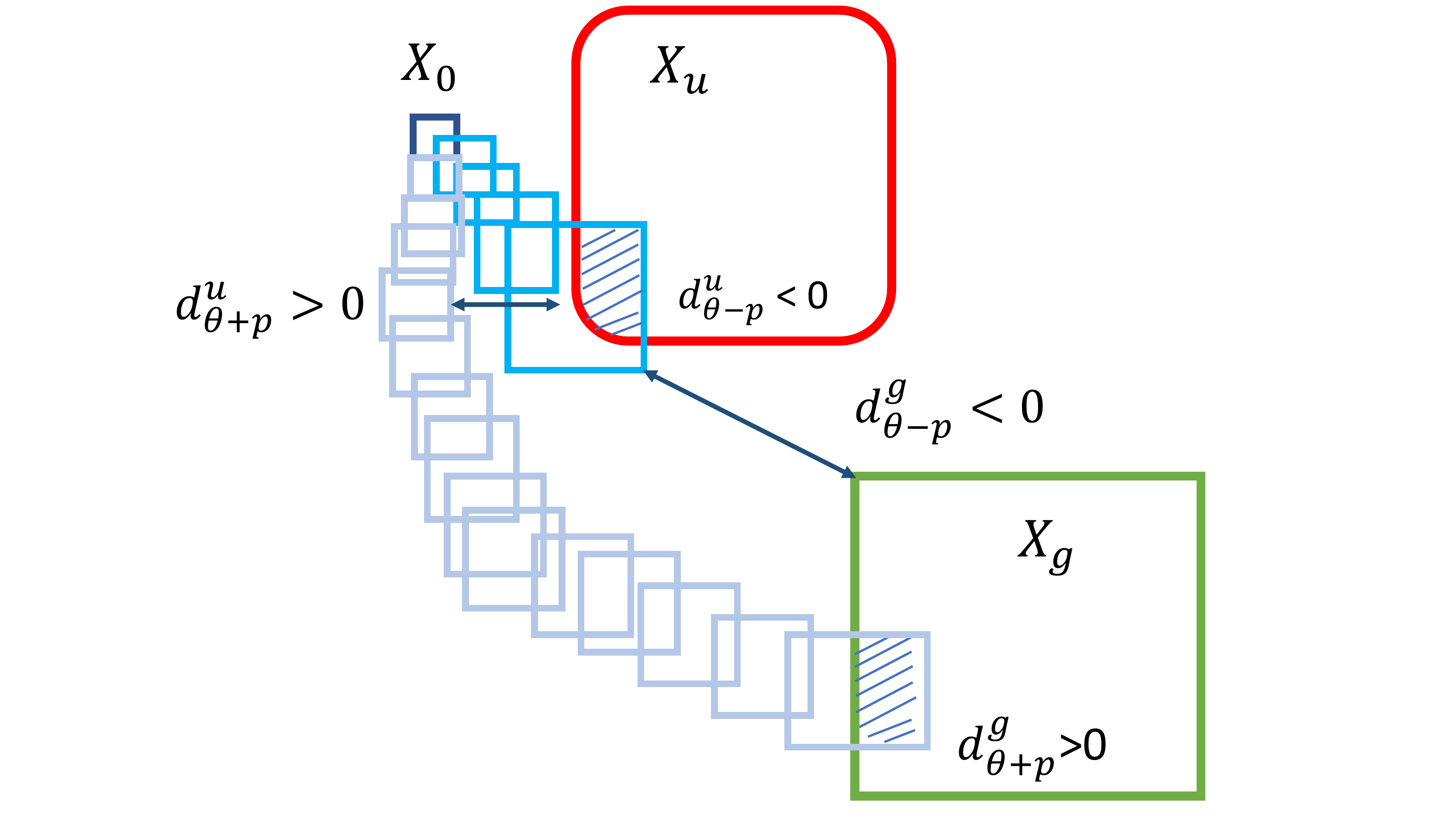}
\caption{
Approximate the gradient for tuning controller parameter by difference method with perturbation $p$.}
\label{fig:difference_method}
\end{minipage}
\end{figure}

We define a geometric distance $d_{\theta}^{u}$ that measures the distance between the reachable set $X_{r}^{T}$ and the unsafe region  $X_u$ as
\begin{equation}
d_{\theta}^{u} = 
\begin{cases}
-|X_{r}^{T}\cap X_{u}|, if \ X_{r}^{T} \cap X_{u} \neq \emptyset \\
 \inf(||x_r - x_u||^{2}), \ \forall x_r \in X_{r}^{T}, \forall x_u \in X_{u}, Otherwise
\end{cases}
\label{eq:heuristic_safe}
\end{equation}

where $|\cdot|$ measures the size of a set. For instance in Fig.~\ref{fig:metrics} with a 2-dimensional system, $|X_{r}^{T}\cap X_{u}|$ is the intersection area between blue and red regions. 
Intuitively, the system is safe within time horizon $T$ if and only if $d_{\theta}^{u}$ is positive. Moreover, the larger the $d_{\theta}^{u}$ is, the further the system stays away from the unsafe region. 

Following the same idea, we define another geometric distance $d_{\theta}^{g}$ for the goal-reaching property as 
\begin{equation}
d_{\theta}^{g} = 
\begin{cases}
\left|X_{r}^{T}\cap X_{g}\right|, if \ X_{r}^{T} \cap X_{g} \neq \emptyset \\
 -\inf(||x_r - x_g||^{2}), \ \forall x_r \in X_{r}^{T}, \forall x_g \in X_{g}, Otherwise
\end{cases}
\label{eq:heuristic_goal}
\end{equation}

The system satisfies the goal-reaching property if and only if $d_{\theta}^{g}$ is positive. Similarly in Fig~\ref{fig:metrics}, the larger the $d_{\theta}^{g}$ is, the better it is for the goal-reaching property. To have formal guarantee on goal-reaching, a searching algorithm for the initial set $X_I \subseteq X_0$  is proposed and detailed later. 

 Based on these two metrics, an optimization problem of controller parameters $\theta$ for the control learning with reach-avoid property can be formulated as
\begin{displaymath}
\begin{cases}
\max_{\theta} \ d_\theta^{u} + \lambda d_\theta^{g}, \\
s.t. \ d_\theta^{u} \geq 0, d_\theta^{g} \geq 0.
\end{cases}
\end{displaymath}
Here $\lambda$ is a hyper-parameter denoting the weight. Overall, a feasible solution $\theta$ of this problem should make both $d_{\theta}^{g}$ and $d_\theta^{u}$ positive, which indicates the reach-avoid property is formally assured.

\paragraph{Wasserstein Distance based Metric:}

Another metric we consider in this paper is Wasserstein distance, which is defined on two distributions $z(x)$ and $v(y)$ as
\begin{equation}\label{eq:Wasserstein}
 W(z, v) = \inf_{\gamma \in \Gamma(x, y)} \int{d(x, y) d\gamma(x, y)},
\end{equation}
where $\Gamma$ denotes the collections of all joint distributions with margins as $z(x)$ and $v(y)$. $d(x, y)$ is a distance measure function over $x, y$, such as norms. 

Our metrics based on the Wasserstein distance are defined as follows. We view the last step of the reachable set $X^{Tl}$ as a uniform distribution $r_{\theta}(x)$, i.e., 
\begin{displaymath}
 r_{\theta}(x) = 
 \begin{cases}
\FS{1}{|X^{Tl}|}, if \ x \in X^{Tl}, \\
 0, otherwise
 \end{cases}
\end{displaymath}
The same applies to the goal set $X_g$ as $g(x)$ and the unsafe set $X_u$ as $u(x)$. With this transformation, Wasserstein distance is naturally defined on $r_{\theta}(x), g(x)$ and $r_{\theta}(x), u(x)$. In this case, the system is reach-avoid if and only if we can determine that $X^T_r \cap X_g \neq \emptyset$ and $X^T_r \cap X_u = \emptyset$. Therefore, the optimization problem based on the Wasserstein distance over controller parameters $\theta$ is defined as
\begin{displaymath}
\begin{cases}
\min_{\theta}  \mathcal{W}(r_{\theta}) = W(r_{\theta}(x), g(x)) - \lambda W(r_{\theta}(x), u(x)) \\
s.t. X^T_r \cap X_g \neq \emptyset, X^T_r \cap X_u = \emptyset
\end{cases}
\end{displaymath}

\subsection{Approximated Gradient Descent Algorithm for Control Learning}
\label{subsection:learning_algorithm}

Based on the computed reachable set and the defined distance metrics, we develop an approximated gradient descent algorithm for the control learning. The framework is shown in Algorithm \ref{alg:Framwork}.

\begin{algorithm}[htb] 
\caption{Verification-in-the-loop Control Learning.} 
\label{alg:Framwork} 
\begin{algorithmic}[1] 
\REQUIRE 

A verifier $\Psi(f, X_0, \kappa_\theta)$ that computes the reachable set $X_{r}^{T}$, system dynamics $f$, initial set $X_0$, initial controller $\kappa_\theta$. \\

\STATE Randomly initialize $\theta$; set the maximum number of updates $N$, control horizon $T$, step lengths $\alpha, \beta$ and $i=0$. 
\WHILE { $i \leq N$ and $X^T_r$ is not reach-avoid}
\STATE Generate perturbation $p$ to get $\theta-p$ and $\theta+p$.

\STATE Compute the reachable sets for each perturbation as $ \Psi(f, X_0, \kappa_{\theta-p})$ and $\Psi(f, X_0, \kappa_{\theta+p})$.

\STATE Compute the geometric distances $\left[d_{\theta-p}^{u}, d_{\theta+p}^{u}, d_{\theta-p}^{g}, d_{\theta+p}^{g}\right]$ with Equations~\eqref{eq:heuristic_safe} and~\eqref{eq:heuristic_goal} or Wasserstein distances $\left[W(r_{\theta+p}, g), W(r_{\theta-p}, g), W(r_{\theta+p}, u), W(r_{\theta-p}, u)\right]$ with Equation~\eqref{eq:Wasserstein}.







\STATE Approximate the gradients $\nabla_{\theta}^{u}$ and $\nabla_{\theta}^{g}$ by Equation~\eqref{eq:gradient}.

\STATE $\theta = \theta - \alpha \nabla_{\theta}^{u} + \beta \nabla_{\theta}^{g}$.

\STATE $i \leftarrow i + 1$.

\ENDWHILE
\STATE Search reach-avoid initial set $X_I$ using Algorithm \ref{alg:searching}.

\STATE \textbf{Return}: Learned controller $\kappa_{\theta}$ and reach-avoid initial set $X_I$.\\
\end{algorithmic}
\end{algorithm}

Because the verifier is often complex and does not have an analytical form, we propose a difference method to approximate the gradients for safety and goal-reaching metrics, as shown in Fig.~\ref{fig:difference_method}. For each update iteration, we generate some perturbations $p$ to the controller $\theta$, and then compute their reachable set and also corresponding metrics.
Thus, for the geometric and Wasserstein metrics, the gradients can be approximated respectively as
\begin{equation}\label{eq:gradient}
\begin{cases}
    \nabla_{\theta}^{u} \approx  \FS{d_{\theta + p}^{u} - d_{\theta - p}^{u} }{2p}, 
    \nabla_{\theta}^{g} \approx  \FS{d_{\theta + p}^{g} - d_{\theta - p}^{g} }{2p}, \\ 
    \nabla_{\theta}^{u} \approx  \FS{W(r_{\theta + p}, u) -  W(r_{\theta - p}, u)}{2p}, 
    \nabla_{\theta}^{g} \approx  \FS{W(r_{\theta + p}, g) -  W(r_{\theta - p}, g)}{2p},    
\end{cases}    
\end{equation}
and thus the controller parameters are updated accordingly. Note that if the reach-avoid property is true for some initial space $X_I \subseteq X_0$, we can directly break from the iteration and return the learned controller. Finally, we search for $X_I$ to complete the algorithm. 



\begin{remark}
An alternate approach is to build a surrogate model such as a neural network to approximate the function of metrics, e.g., $d_\theta^u$ over controller parameters $\theta$ in a black-box manner, and directly obtain the gradients from such a model with current tools such as Tensorflow and Pytorch.  
\end{remark}

\paragraph{Reach-avoid Initial Set Searching:} Once Algorithm~\ref{alg:Framwork} successfully learns a controller,
safety can be ensured to the entire initial set $X_0$. However, goal-reaching is not guaranteed for $X_0$ because of the intersection operator we used in the metrics and also due to the over-approximation computation of reachable set. Thus, we further propose a searching algorithm to obtain the reach-avoid initial set $X_I \subseteq X_0$ such that $\forall x(0) \in X_I$, the reach-avoid property is formally verified to hold.

\begin{algorithm}[htb] 
\caption{$X_I$ Searching.} 
\label{alg:searching}
\begin{algorithmic}[1] 
\REQUIRE 
Verifier $\Psi(f, X_0, \kappa_\theta)$, system dynamics $f$, initial set $X_0$, learned controller $\kappa_\theta$, $X_I =\emptyset$.\\

\WHILE{$X_I$ not converged}
\STATE Evenly partition $X_0$ into $P$ sub-spaces $X_p(p=1, \cdots, P)$. 
\FOR{$p \in (1, \cdots, P)$}
\IF {$\Psi(f, X_p, \kappa_{\theta})$ is reach-avoid}

\STATE $X_I = X_I \cup X_p$, $X_0 = X_0 - X_p$.

\ENDIF
\ENDFOR
\STATE $P *= 2$.
\ENDWHILE
\end{algorithmic}
\end{algorithm}

\paragraph{Optimality Analysis:}
\label{sec:optimality}
 
The geometric distances $d_\theta^{u}, d_\theta^{g} $ are not convex functions over the reachable set $X_r^T$. The Wasserstein distance $\mathcal{W}(r_{\theta})$ is believed to be convex and almost everywhere differentiable in the distribution $r_{\theta}$~\cite{arjovsky2017wasserstein, rabin2015convex, peyre2019computational}. Due to its convexity, it holds that

\begin{displaymath}
\mathcal{W}(r_{\theta_1}) - \mathcal{W}(r_{\theta_2}) \leq \left<\delta \mathcal{W} / \delta r_{\theta_1}, r_{\theta_1} - r_{\theta_2} \right>, \forall r_{\theta_1}, r_{\theta_2} \in \mathcal{R},
\end{displaymath}
where $\left<\cdot , \cdot \right>$ is the inner-product. 

Let $\hat{\theta}$ be an $\epsilon$-stationary point of objective $\mathcal{W}(r_\theta)$. 
In most cases of the experiments, our approach can reach such stationary points.
It then holds with some constant number $M$ that
\begin{displaymath}
\nabla_\theta \mathcal{W}(r_{\hat{\theta}})^T v \leq \epsilon, \forall v \in \mathcal{B} = \left\{ \theta \in \mathbb{R}^n , ||\theta||^2 \leq  M \right\}
\end{displaymath}

We assume that $r_{\theta}$ is a differentiable function of $\theta$. 
By the chain rule, it then holds that
\begin{displaymath}
\nabla_\theta \mathcal{W}(r_{\hat{\theta}})^T v = \left < \delta \mathcal{W} / \delta r_{\hat{\theta}}, \left(\at{\frac{dr_{\theta}}{d\theta}}{\theta=\hat{\theta}} \right)^{T} v \right> \leq \epsilon, \forall v \in \mathcal{B}
\end{displaymath}

Let $r^{*}$ be a global minimizer of $\mathcal{W}(r)$ for $r\in  \mathcal{R}$ with corresponding $\theta^{*}$, we then have 
\begin{displaymath}
\mathcal{W}(r_{\hat{\theta}}) - \mathcal{W}(r^*) \leq \left<\delta \mathcal{W} / \delta r_{\hat{\theta}}, r_{\hat{\theta}} - r^{*} \right>
\end{displaymath}

Let $\varphi_{\hat{\theta}} = \at{\frac{dr_{\theta}}{d\theta}}{\theta=\hat{\theta}}$ represents the tangent function of $r_{\theta}$ at $\theta = \hat{\theta}$. Then, by combining the above two equations, we have
\begin{displaymath}
\begin{aligned}
  \mathcal{W}(r_{\hat{\theta}}) - \mathcal{W}(r^*) &\leq \epsilon + \left<\delta \mathcal{W} / \delta r_{\hat{\theta}}, r_{\hat{\theta}} - r^{*} - \varphi_{\hat{\theta}}^T v \right>, \forall v \in \mathcal{B} \\ &\leq \epsilon + ||\delta \mathcal{W} / \delta r_{\hat{\theta}}|| \cdot ||r_{\hat{\theta}} - r^{*} - \varphi_{\hat{\theta}}^T v||, \forall v \in \mathcal{B} 
\end{aligned}
\end{displaymath}

By taking the infimum on the right hand side over $v \in \mathcal{B}$, we can now obtain the optimality bound of the stationary point to the global optimum as 
\begin{displaymath}
\mathcal{W}(r_{\hat{\theta}}) - \mathcal{W}(r^*) \leq \epsilon + ||\delta \mathcal{W} / \delta r_{\hat{\theta}}|| \cdot \inf_{v\in \mathcal{B}} ||r_{\hat{\theta}} - r^{*} - \varphi_{\hat{\theta}}^T v|| 
\end{displaymath}

This shows that on the Wasserstein distance metrics, our approach is highly likely to reach a stationary point that has a bounded distance to the global optimum.



\section{Experimental Results}
\label{sec:experiment}

\noindent
\textbf{Test systems:} We evaluate our approach by learning linear controllers for a linear adaptive cruise control (ACC) system and neural network controllers for a Van der Pol's oscillator system. The Baselines include LQR~\cite{wang2020energy} controller and deep deterministic policy gradient(DDPG) method~\cite{lillicrap2016continuous}.



\begin{wrapfigure}{r}{0.4\textwidth}
  \begin{center}
    \includegraphics[width=0.8\linewidth]{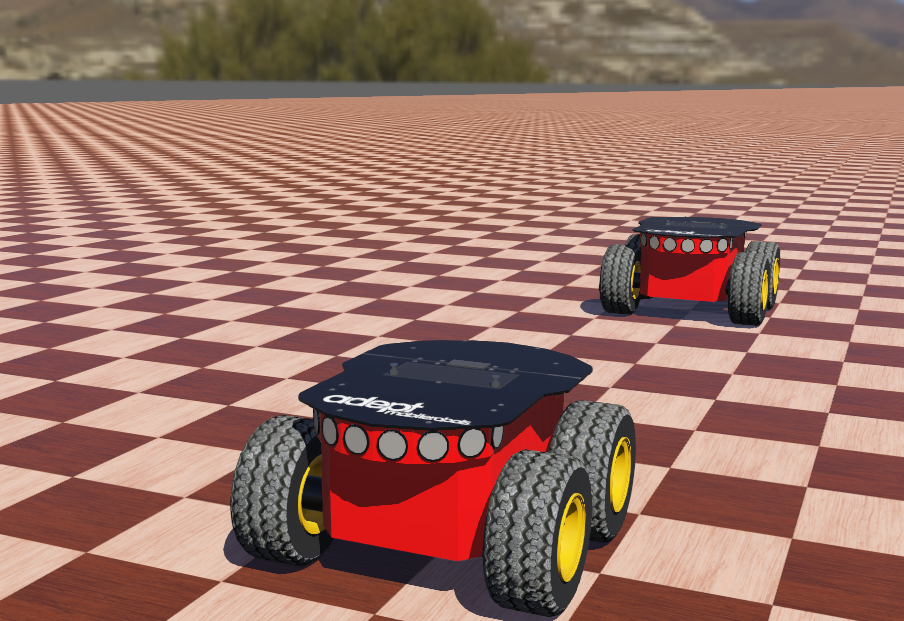}
  \end{center}
  \caption{Four-wheeled robot "Pioneer 3-AT" environment for ACC in Webots.}
  \label{fig:webots}
\end{wrapfigure}

\textbf{ACC:} There are two robotic vehicles driving on the road, shown in Fig.~\ref{fig:webots} with the Webots environment~\cite{Webots}. 
The front vehicle drives at a velocity $v_f$ while the ego vehicle manages the relative distance between the two vehicles by accelerating or braking. The dynamics of the ACC system can be expressed as
$\dot{s} = v_f - v,
\dot{v} = kv + u,$
where $v_f = 40$ is the velocity of the front vehicle, $k = -0.2$ is the resistance related to the velocity of the ego vehicle, and $\delta = 0.1$ is the sampling period. $(s, v)$ is the system state vector where $s$ is the relative distance and $v$ is the ego car's velocity. The system has initial space $X_0 = \{ (s, v) \ | \ s \in [122, 124], v \in [48, 52] \}$, the unsafe set $X_u  = \{ (s, v) \ | \ s \leq 120 \}$ and the goal set $X_g = \{ (s, v) \ | \ s \in [145, 155], v \in [39.5, 40.5] \}$. 
Initially, $v \in [48, 52] > 40$, and thus the distance $s$ is being reduced and the system is approaching the unsafe region. 
 
\begin{figure}[htbp]
\centering
\begin{minipage}[t]{0.48\textwidth}
\centering
\includegraphics[width=1.05\linewidth]{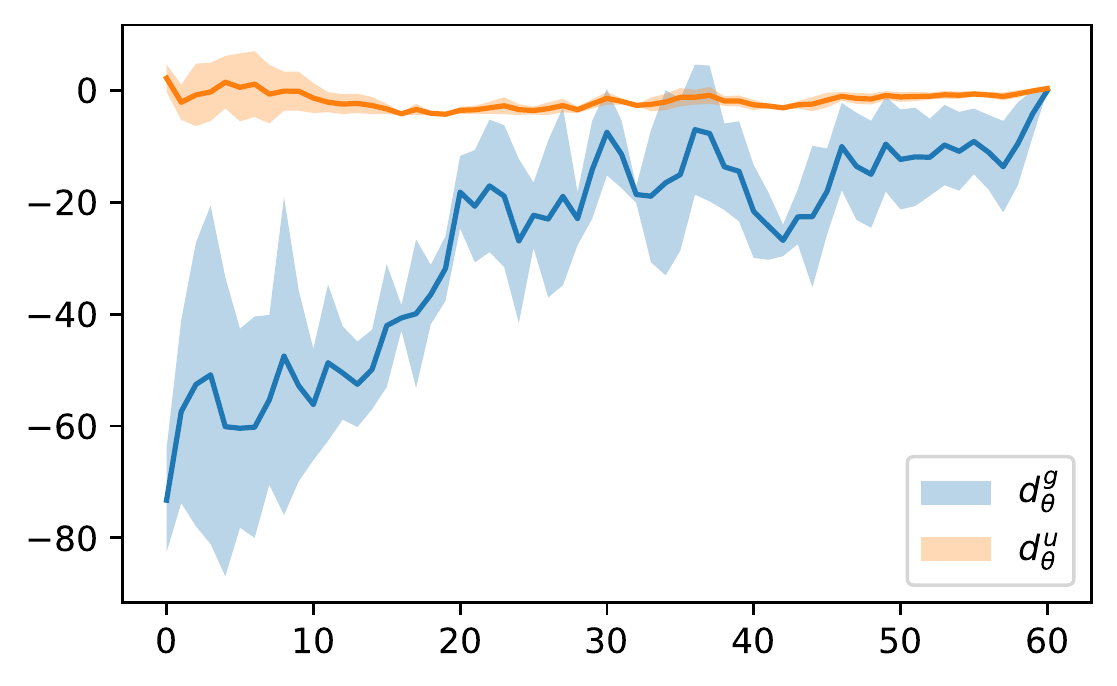}
    \caption{Learning curves for the geometric distance $d_\theta^{g}$ and $d_\theta^{u}$ in ACC. The algorithm successfully learned a reach-avoid controller with around 60 iterations with final $d_{\theta}^{g}$ and $d_{\theta}^{u}$ being positive. 
     The shaded region is the standard deviation for 5 individual runs.}
    
    \label{fig:ACC_learning}
\end{minipage}
\quad
\begin{minipage}[t]{0.48\textwidth}
\centering
\includegraphics[width=1.1\linewidth]{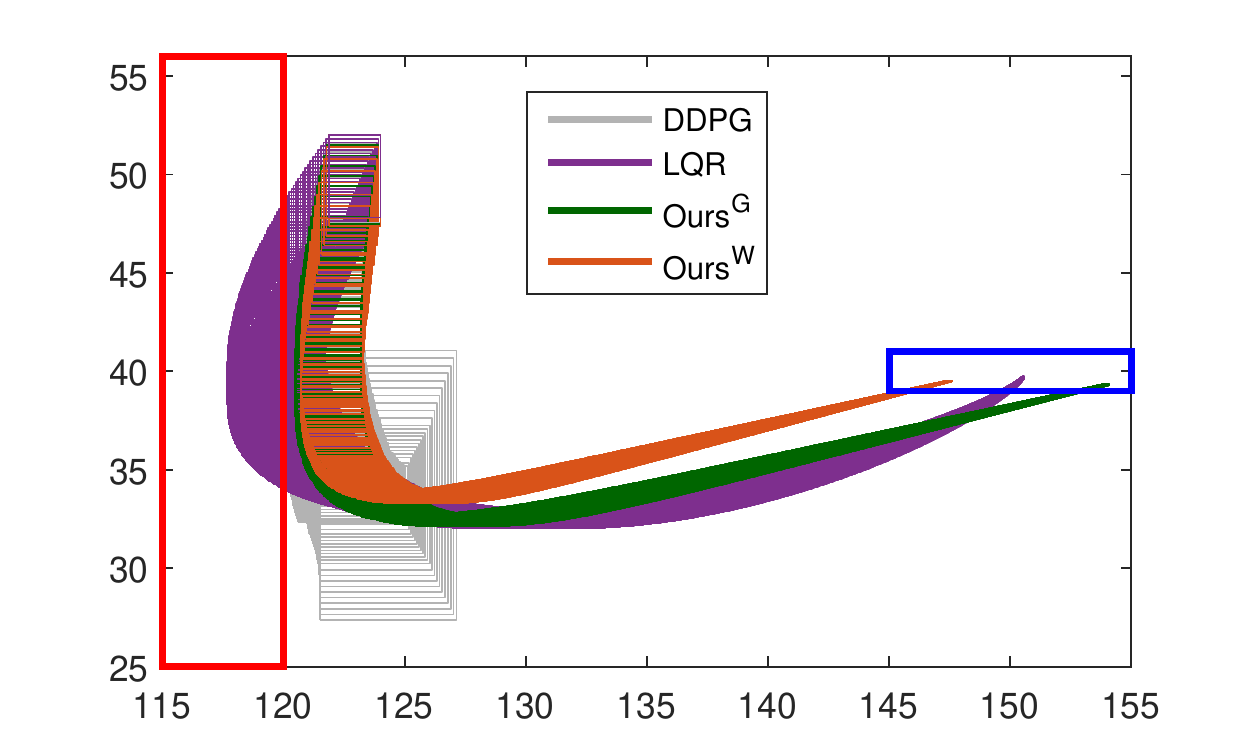}
    \caption{Reachable sets of our approach with geometric and Wasserstein distance and of Baselines for ACC. Goal region is within the blue boundary and the unsafe region is within red. Our controllers 
    are verified to be reach-avoid with $X_I = X_0$ while DDPG and LQR cannot.}
\label{fig:ACC_reachable}
\end{minipage}
\end{figure}

 \noindent
 \textit{Results Comparison:} The comparison between different approaches on the ACC example is shown in Table~\ref{table:ACC}. 
A typical learning curve is shown in Fig.~\ref{fig:ACC_learning}. With verification in the loop, our approaches on the Wasserstein metric, denoted as Ours(W), and on the geometric metric, denoted as Ours(G), show much faster convergence rate (around 64 and 62 updates, respectively) than the DDPG method (more than 13K updates, defined on a threshold of the total reward). 

 Considering \emph{experimental} safety and goal-reaching properties, we discretized the system with zeroth-order hold and simulated the system traces with 500 randomly picked initial states $x(0)$ from $X_0$.  Our approach achieves both $100\%$ safe control rate and goal-reaching rate, while DDPG and LQR\cite{wang2020energy} cannot. As shown in Fig.~\ref{fig:ACC_reachable}, controllers from our approach are \emph{formally verified to satisfy the reach-avoid property}, while DDPG failed due to the explosion of the reachable set computation after around 10 steps and the LQR controller is unsafe.

\begin{table}
\centering
  \caption{Comparison of results for the ACC experiment. The controllers synthesized from our verification-in-the-loop approach are guaranteed to be reach-avoid while DDPG's is unknown due to the over-approximation of its reachable set computation. With zeroth-order hold, our controllers achieve $100\%$ safe control rate and goal-reaching rate with the discretized simulation while DDPG and LQR do not. The learning process converges much faster with our approach than using DDPG.}
  \begin{tabular}{ccccl}
    \hline
     & Convergence iterations & Safe control rate & Goal-reaching rate & Verified result\\
    \hline
    LQR\cite{wang2020energy} & - & 73\% & 27\% & Unsafe\\
    DDPG & $13.6(\pm 2.1)$K & 99.8 \% & 99.8\%   & Unknown \\
        Ours(W) & $64(\pm31.6)$ &  100\% & 100\% &  reach-avoid\\
    Ours(G) & $62(\pm 6.1)$ & 100 \% & 100 \% & reach-avoid \\
  \hline
  \label{table:ACC}
\end{tabular}
\end{table}


\textbf{Oscillator:}
Van der Pol's oscillator is a 2-dimensional non-linear benchmark system, which can be expressed as 
\begin{displaymath}
\begin{cases}
\dot{x_1} = x_2,\\
\dot{x_2} = \gamma(1- x_1^2)x_2 - x_1 + u,
\end{cases}
\end{displaymath}
where $(x_1, x_2)$ is the system state, sampling period $\delta=0.1$, and damping coefficient $\gamma = 1$. The initial set $X_0$ is $[-0.51, -0.49] \times [0.49, 0.51]$. The goal set for this example is $X_g = [-0.05, 0.05] \times  [-0.05, 0.05]$, and the unsafe region $X_u = [-0.3, -0.25] \times [0.2, 0.35]$. 


\smallskip
\begin{table}
\centering
  \caption{Comparison of results for the Oscillator experiment. Our approach again outperforms DDPG for every comparison metric. }
  \label{table:OS}
  \begin{tabular}{ccccl}
    \hline
     & Convergence iterations & Safe control rate & Goal-reaching rate &  Verified result\\
    \hline
    DDPG & $13.7(\pm 6.2)$K & 100 \% & 79.2 \%  & Unknown \\
    Ours(W) & $55(\pm13)$ & 100\% & 100\% & reach-avoid \\
    Ours(H) & $65(\pm4)$ & 100 \% & 100 \% & reach-avoid \\
  \hline
\end{tabular}
\end{table}

\begin{figure}[tbp]
\centering
\begin{minipage}[t]{0.48\textwidth}
\centering
    \includegraphics[width=1.0\linewidth]{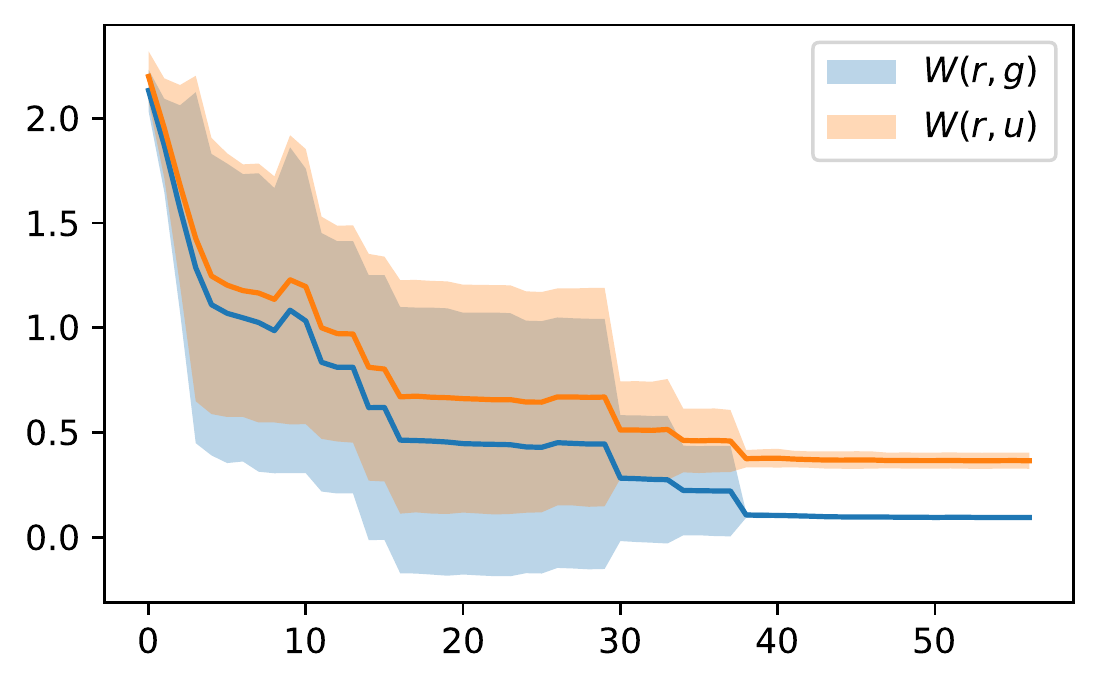}
    \caption{Learning curves for the Wasserstein distances $W(r, g)$ and $W(r, u)$ for the oscillator system. The algorithm successfully learned a controller with around 56 updates. The shaded region is the standard deviation for 5 individual runs.}
    \label{fig:OS_learning}
\end{minipage}
\quad
\begin{minipage}[t]{0.48\textwidth}
\centering
\includegraphics[width=1.1\linewidth]{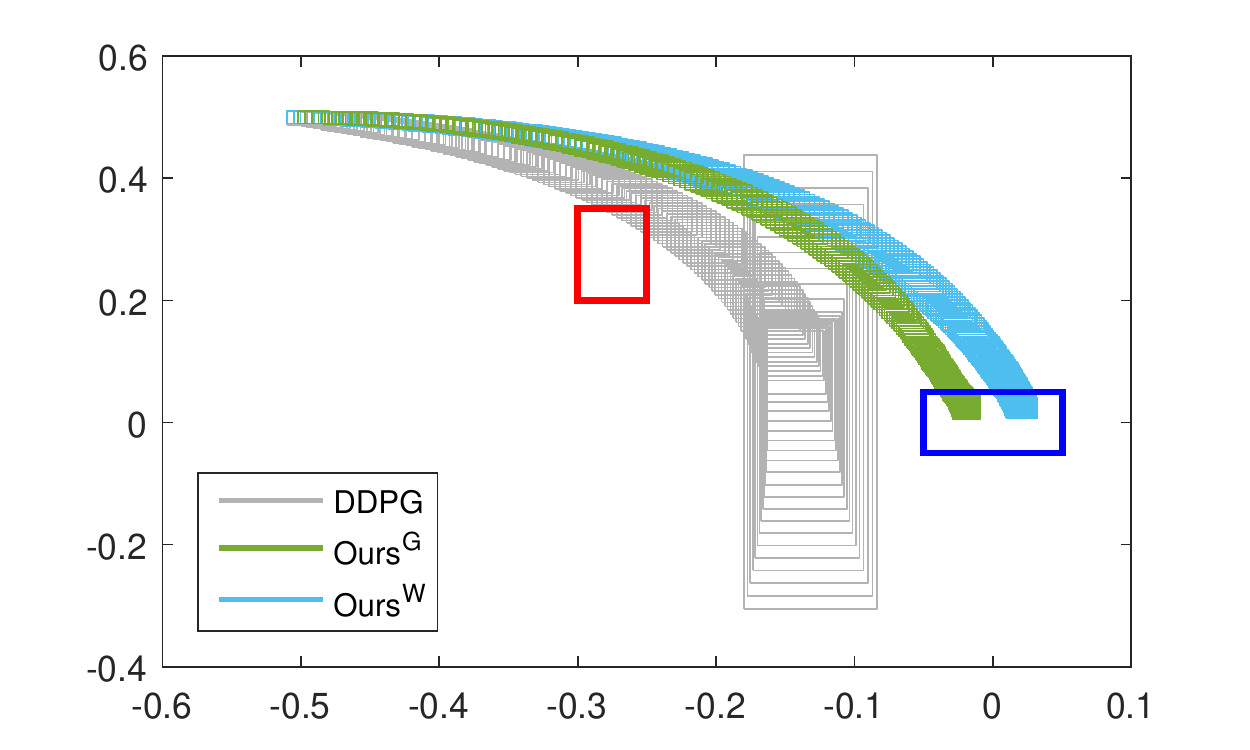}
     \caption{DDPG controller is verified to be unknown due to the over-approximation of reachable set. The learned neural network controllers from our approach are reach-avoid with $X_I = [-0.502, -0.49] \times [0.49, 0.51](G)$ and $X_I = X_0 = [-0.51, -0.49] \times [0.49, 0.51](W)$. }
     \label{fig:OS_reachable}
\end{minipage}
\end{figure}

\noindent
 \textit{Results Comparison:} The comparison between different methods is shown in Table~\ref{table:OS}, 
 and a typical learning curve for this system is shown in Fig.~\ref{fig:OS_learning}. Our approach converges much faster than the DDPG method. We also discretized the oscillator with zeroth-order hold and simulated the traces from 500 random initial states $x(0)$ from $X_0$. Our approach achieves  $100\%$ experimental safe control and goal-reaching rates for both controllers (synthesized using Wasserstein or geometric metrics), while DDPG's controller failed. As shown in Fig.~\ref{fig:OS_reachable}, controllers from our approach are formally guaranteed to be reach-avoid while DDPG is not. 



 
 
\section{Discussions}\label{sec:discussion}
\textbf{Hybrid Control Learning:} To achieve reach-avoid property, ACC naturally requires the controller to first brake to avoid the unsafe region and then accelerate to reach the goal region. In such case, ACC requires the controller to have a relatively big Lipschitz constant. However, the reachable set computation with ReachNN may encounter large over-approximation errors when addressing a neural network controller with a big Lipschitz constant. In this case, the parameter perturbation may not cause any change on the reachable set and our approach could fail to learn a controller. To mitigate this issue, we propose a hybrid controller learning method. Specifically, by decomposing the reach-avoid learning problem into two sub-problems, we learn a neural network controller with few reachable steps to avoid the unsafe region and then a linear controller to make the system approach the goal set. Please see the \textbf{Appendix} for more details.  


\textbf{Verification Tightness:} Similarly as other verification tools~\cite{ivanov2019verisig, dutta2019reachability}, ReachNN computes an over-approximation of the reachable set. The tightness of such over-approximation has significant impact on our verification-in-the-loop approach. There are adjustable parameters for changing the tightness in ReachNN~\cite{huang2019reachnn} (and also other tools~\cite{ivanov2019verisig, dutta2019reachability, chen2013flow}). Intuitively, tighter verification consumes more computation resources and takes more time to finish. 
For Wasserstein distance on the oscillator system, the tighter reachable set computation in average takes around 40 steps with near 115 seconds for each step to learn a neural network controller, compared to the lesser tight computation that takes about 55 iterations with 86 seconds for each step. Please see the \textbf{Appendix} for more details.  
\section{Conclusion}\label{sec:conclusion}
In this paper, we propose a correct-by-construction control learning framework with reach-avoid guarantees by integrating the verification in a closed-loop manner. Our approach first constructs the control feedback metrics with the reachable sets computed by the verifier, and then iteratively tunes the controller parameters with approximated gradients until a feasible solution is found. Experiments on linear and non-linear systems with linear and neural network controllers demonstrate the effectiveness of our approach on convergence iterations, safe/goal-reaching rate, and verified results.

\bibliographystyle{unsrtnat}
\bibliography{yixuan_bib}

\begin{thebibliography}{36}
\providecommand{\natexlab}[1]{#1}
\providecommand{\url}[1]{\texttt{#1}}
\expandafter\ifx\csname urlstyle\endcsname\relax
  \providecommand{\doi}[1]{doi: #1}\else
  \providecommand{\doi}{doi: \begingroup \urlstyle{rm}\Url}\fi

\bibitem[Cofer et~al.(2020)Cofer, Amundson, Sattigeri, Passi, Boggs, Smith,
  Gilham, Byun, and Rayadurgam]{cofer2020run}
Darren Cofer, Isaac Amundson, Ramachandra Sattigeri, Arjun Passi, Christopher
  Boggs, Eric Smith, Limei Gilham, Taejoon Byun, and Sanjai Rayadurgam.
\newblock Run-time assurance for learning-based aircraft taxiing.
\newblock In \emph{2020 AIAA/IEEE 39th Digital Avionics Systems Conference
  (DASC)}, pages 1--9. IEEE, 2020.

\bibitem[Koopman and Wagner(2017)]{koopman2017autonomous}
Philip Koopman and Michael Wagner.
\newblock Autonomous vehicle safety: An interdisciplinary challenge.
\newblock \emph{IEEE Intelligent Transportation Systems Magazine}, 9\penalty0
  (1):\penalty0 90--96, 2017.

\bibitem[Sutton and Barto(2018)]{sutton2018reinforcement}
Richard~S Sutton and Andrew~G Barto.
\newblock \emph{Reinforcement learning: An introduction}.
\newblock MIT press, 2018.

\bibitem[Wang et~al.(2021{\natexlab{a}})Wang, Huang, Wang, Xu, Wang, and
  Zhu]{wang2021cocktail}
Yixuan Wang, Chao Huang, Zhilu Wang, Shichao Xu, Zhaoran Wang, and Qi~Zhu.
\newblock Cocktail: Learn a better neural network controller from multiple
  experts via adaptive mixing and robust distillation.
\newblock \emph{arXiv preprint arXiv:2103.05046}, 2021{\natexlab{a}}.

\bibitem[Ivanov et~al.(2020)Ivanov, Carpenter, Weimer, Alur, Pappas, and
  Lee]{ivanov2020case}
Radoslav Ivanov, Taylor~J Carpenter, James Weimer, Rajeev Alur, George~J
  Pappas, and Insup Lee.
\newblock Case study: verifying the safety of an autonomous racing car with a
  neural network controller.
\newblock In \emph{Proceedings of the 23rd International Conference on Hybrid
  Systems: Computation and Control}, pages 1--7, 2020.

\bibitem[Margellos and Lygeros(2011)]{margellos2011hamilton}
Kostas Margellos and John Lygeros.
\newblock Hamilton--jacobi formulation for reach--avoid differential games.
\newblock \emph{IEEE Transactions on Automatic Control}, 56\penalty0
  (8):\penalty0 1849--1861, 2011.

\bibitem[Jin et~al.(2020)Jin, Wang, Yang, and Mou]{jin2020neural}
Wanxin Jin, Zhaoran Wang, Zhuoran Yang, and Shaoshuai Mou.
\newblock Neural certificates for safe control policies.
\newblock \emph{arXiv preprint arXiv:2006.08465}, 2020.

\bibitem[Blondel and Tsitsiklis(2000)]{blondel2000survey}
Vincent~D Blondel and John~N Tsitsiklis.
\newblock A survey of computational complexity results in systems and control.
\newblock \emph{Automatica}, 36\penalty0 (9):\penalty0 1249--1274, 2000.

\bibitem[Henzinger et~al.(1998)Henzinger, Kopke, Puri, and
  Varaiya]{henzinger1998s}
Thomas~A Henzinger, Peter~W Kopke, Anuj Puri, and Pravin Varaiya.
\newblock What's decidable about hybrid automata?
\newblock \emph{Journal of computer and system sciences}, 57\penalty0
  (1):\penalty0 94--124, 1998.

\bibitem[Xiang and Johnson(2018)]{xiang2018reachability}
Weiming Xiang and Taylor~T Johnson.
\newblock Reachability analysis and safety verification for neural network
  control systems.
\newblock \emph{arXiv preprint arXiv:1805.09944}, 2018.

\bibitem[Aydinoglu et~al.(2020)Aydinoglu, Fazlyab, Morari, and
  Posa]{aydinoglu2020stability}
Alp Aydinoglu, Mahyar Fazlyab, Manfred Morari, and Michael Posa.
\newblock Stability analysis of complementarity systems with neural network
  controllers.
\newblock \emph{arXiv preprint arXiv:2011.07626}, 2020.

\bibitem[Agachi et~al.(2016)Agachi, Cristea, Csavdari, and
  Szilagyi]{agachi20162}
Paul~Serban Agachi, Mircea~Vasile Cristea, Alexandra~Ana Csavdari, and Botond
  Szilagyi.
\newblock \emph{2. Model predictive control}.
\newblock De Gruyter, 2016.

\bibitem[Bemporad et~al.(2002)Bemporad, Morari, Dua, and
  Pistikopoulos]{bemporad2002explicit}
Alberto Bemporad, Manfred Morari, Vivek Dua, and Efstratios~N Pistikopoulos.
\newblock The explicit linear quadratic regulator for constrained systems.
\newblock \emph{Automatica}, 38\penalty0 (1):\penalty0 3--20, 2002.

\bibitem[Lillicrap et~al.(2016)Lillicrap, Hunt, Pritzel, Heess, Erez, Tassa,
  Silver, and Wierstra]{lillicrap2016continuous}
Timothy~P Lillicrap, Jonathan~J Hunt, Alexander Pritzel, Nicolas Heess, Tom
  Erez, Yuval Tassa, David Silver, and Daan Wierstra.
\newblock Continuous control with deep reinforcement learning.
\newblock In \emph{ICLR (Poster)}, 2016.

\bibitem[Schulman et~al.(2017)Schulman, Wolski, Dhariwal, Radford, and
  Klimov]{schulman2017proximal}
John Schulman, Filip Wolski, Prafulla Dhariwal, Alec Radford, and Oleg Klimov.
\newblock Proximal policy optimization algorithms.
\newblock \emph{arXiv preprint arXiv:1707.06347}, 2017.

\bibitem[Ivanov et~al.(2019)Ivanov, Weimer, Alur, Pappas, and
  Lee]{ivanov2019verisig}
Radoslav Ivanov, James Weimer, Rajeev Alur, George~J Pappas, and Insup Lee.
\newblock Verisig: verifying safety properties of hybrid systems with neural
  network controllers.
\newblock In \emph{Proceedings of the 22nd ACM International Conference on
  Hybrid Systems: Computation and Control}, pages 169--178, 2019.

\bibitem[Huang et~al.(2019)Huang, Fan, Li, Chen, and Zhu]{huang2019reachnn}
Chao Huang, Jiameng Fan, Wenchao Li, Xin Chen, and Qi~Zhu.
\newblock Reachnn: Reachability analysis of neural-network controlled systems.
\newblock \emph{ACM Transactions on Embedded Computing Systems (TECS)},
  18\penalty0 (5s):\penalty0 1--22, 2019.

\bibitem[Chen et~al.(2013)Chen, {\'A}brah{\'a}m, and
  Sankaranarayanan]{chen2013flow}
Xin Chen, Erika {\'A}brah{\'a}m, and Sriram Sankaranarayanan.
\newblock Flow*: An analyzer for non-linear hybrid systems.
\newblock In \emph{International Conference on Computer Aided Verification},
  pages 258--263. Springer, 2013.

\bibitem[Henderson et~al.(2018)Henderson, Islam, Bachman, Pineau, Precup, and
  Meger]{henderson2018deep}
Peter Henderson, Riashat Islam, Philip Bachman, Joelle Pineau, Doina Precup,
  and David Meger.
\newblock Deep reinforcement learning that matters.
\newblock In \emph{Proceedings of the AAAI Conference on Artificial
  Intelligence}, volume~32, 2018.

\bibitem[Dutta et~al.(2019)Dutta, Chen, and
  Sankaranarayanan]{dutta2019reachability}
Souradeep Dutta, Xin Chen, and Sriram Sankaranarayanan.
\newblock Reachability analysis for neural feedback systems using regressive
  polynomial rule inference.
\newblock In \emph{Proceedings of the 22nd ACM International Conference on
  Hybrid Systems: Computation and Control}, pages 157--168, 2019.

\bibitem[Dreossi et~al.(2019{\natexlab{a}})Dreossi, Donz{\'e}, and
  Seshia]{dreossi2019compositional}
Tommaso Dreossi, Alexandre Donz{\'e}, and Sanjit~A Seshia.
\newblock Compositional falsification of cyber-physical systems with machine
  learning components.
\newblock \emph{Journal of Automated Reasoning}, 63\penalty0 (4):\penalty0
  1031--1053, 2019{\natexlab{a}}.

\bibitem[Fremont et~al.(2020)Fremont, Chiu, Margineantu, Osipychev, and
  Seshia]{fremont2020formal}
Daniel~J Fremont, Johnathan Chiu, Dragos~D Margineantu, Denis Osipychev, and
  Sanjit~A Seshia.
\newblock Formal analysis and redesign of a neural network-based aircraft
  taxiing system with verifai.
\newblock In \emph{International Conference on Computer Aided Verification},
  pages 122--134. Springer, 2020.

\bibitem[Dreossi et~al.(2019{\natexlab{b}})Dreossi, Fremont, Ghosh, Kim,
  Ravanbakhsh, Vazquez-Chanlatte, and Seshia]{dreossi2019verifai}
Tommaso Dreossi, Daniel~J Fremont, Shromona Ghosh, Edward Kim, Hadi
  Ravanbakhsh, Marcell Vazquez-Chanlatte, and Sanjit~A Seshia.
\newblock Verifai: A toolkit for the formal design and analysis of artificial
  intelligence-based systems.
\newblock In \emph{International Conference on Computer Aided Verification},
  pages 432--442. Springer, 2019{\natexlab{b}}.

\bibitem[Qin et~al.(2021)Qin, Zhang, Chen, Chen, and Fan]{qin2021learning}
Zengyi Qin, Kaiqing Zhang, Yuxiao Chen, Jingkai Chen, and Chuchu Fan.
\newblock Learning safe multi-agent control with decentralized neural barrier
  certificates.
\newblock \emph{arXiv preprint arXiv:2101.05436}, 2021.

\bibitem[Zhao et~al.(2020)Zhao, Zeng, Chen, and Liu]{zhao2020synthesizing}
Hengjun Zhao, Xia Zeng, Taolue Chen, and Zhiming Liu.
\newblock Synthesizing barrier certificates using neural networks.
\newblock In \emph{Proceedings of the 23rd International Conference on Hybrid
  Systems: Computation and Control}, pages 1--11, 2020.

\bibitem[Chang et~al.(2020)Chang, Roohi, and Gao]{chang2020neural}
Ya-Chien Chang, Nima Roohi, and Sicun Gao.
\newblock Neural lyapunov control.
\newblock \emph{arXiv preprint arXiv:2005.00611}, 2020.

\bibitem[Garc{\i}a and Fern{\'a}ndez(2015)]{garcia2015comprehensive}
Javier Garc{\i}a and Fernando Fern{\'a}ndez.
\newblock A comprehensive survey on safe reinforcement learning.
\newblock \emph{Journal of Machine Learning Research}, 16\penalty0
  (1):\penalty0 1437--1480, 2015.

\bibitem[Fulton and Platzer(2018)]{fulton2018safe}
Nathan Fulton and Andr{\'e} Platzer.
\newblock Safe reinforcement learning via formal methods: Toward safe control
  through proof and learning.
\newblock In \emph{Proceedings of the AAAI Conference on Artificial
  Intelligence}, volume~32, 2018.

\bibitem[Alshiekh et~al.(2018)Alshiekh, Bloem, Ehlers, K{\"o}nighofer, Niekum,
  and Topcu]{alshiekh2018safe}
Mohammed Alshiekh, Roderick Bloem, R{\"u}diger Ehlers, Bettina K{\"o}nighofer,
  Scott Niekum, and Ufuk Topcu.
\newblock Safe reinforcement learning via shielding.
\newblock In \emph{Proceedings of the AAAI Conference on Artificial
  Intelligence}, volume~32, 2018.

\bibitem[Li and Bastani(2020)]{li2020robust}
Shuo Li and Osbert Bastani.
\newblock Robust model predictive shielding for safe reinforcement learning
  with stochastic dynamics.
\newblock In \emph{2020 IEEE International Conference on Robotics and
  Automation (ICRA)}, pages 7166--7172. IEEE, 2020.

\bibitem[Wang et~al.(2021{\natexlab{b}})Wang, Huang, Wang, Hobbs, Chakraborty,
  and Zhu]{wang_date_21}
Zhilu Wang, Chao Huang, Yixuan Wang, Clara Hobbs, Samarjit Chakraborty, and
  Qi~Zhu.
\newblock Bounding perception neural network uncertainty for safe control of
  autonomous systems.
\newblock \emph{Design, Automation and Test in Europe Conference and Exhibition
  (DATE)}, 2021{\natexlab{b}}.

\bibitem[Arjovsky et~al.(2017)Arjovsky, Chintala, and
  Bottou]{arjovsky2017wasserstein}
Martin Arjovsky, Soumith Chintala, and L{\'e}on Bottou.
\newblock Wasserstein generative adversarial networks.
\newblock In \emph{International conference on machine learning}, pages
  214--223. PMLR, 2017.

\bibitem[Rabin and Papadakis(2015)]{rabin2015convex}
Julien Rabin and Nicolas Papadakis.
\newblock Convex color image segmentation with optimal transport distances.
\newblock In \emph{International conference on scale space and variational
  methods in computer vision}, pages 256--269. Springer, 2015.

\bibitem[Peyr{\'e} et~al.(2019)Peyr{\'e}, Cuturi,
  et~al.]{peyre2019computational}
Gabriel Peyr{\'e}, Marco Cuturi, et~al.
\newblock Computational optimal transport: With applications to data science.
\newblock \emph{Foundations and Trends{\textregistered} in Machine Learning},
  11\penalty0 (5-6):\penalty0 355--607, 2019.

\bibitem[Wang et~al.(2020)Wang, Huang, and Zhu]{wang2020energy}
Yixuan Wang, Chao Huang, and Qi~Zhu.
\newblock Energy-efficient control adaptation with safety guarantees for
  learning-enabled cyber-physical systems.
\newblock In \emph{2020 IEEE/ACM International Conference On Computer Aided
  Design (ICCAD)}, pages 1--9. IEEE, 2020.

\bibitem[Webots()]{Webots}
Webots.
\newblock http://www.cyberbotics.com.
\newblock URL \url{http://www.cyberbotics.com}.
\newblock Open-source Mobile Robot Simulation Software.

\end{thebibliography}

\section{Appendix}

\subsection{Additional Experimental Details}

The main experimental results are presented in the Section 4 of the main paper. Here we provide additional experimental details for the ACC and oscillator systems, including the learning hyper-parameters, learned controllers, baseline settings, etc.  
The experiments are conducted on a i7-CPU laptop with 16GB main memory.

\textbf{ACC:} We learn a linear controller that is expressed as 
\begin{displaymath}
\kappa_\theta((s, v)) = \theta_1 s + \theta_2 v,
\end{displaymath}
where $\theta = (\theta_1, \theta_2)$ is initialized with random variables from $[0, 1]^2$. 
 The reachable set horizon is set to 50 steps. The step length of the gradient descent is set to $10^{-4}$ for both safety $d_\theta^{u}$ and goal-reaching $d_\theta^{g}$ geometric distances, as well as for $W(r, g)$ and $W(r, u)$ Wasserstein distances.  The perturbations $p$ on the controller parameters are randomly and uniformly sampled from $[-1, 1]$. The baseline DDPG method fails to learn a linear controller in the experiments, so we design a very small neural network controller with 1 hidden layer of 10 neurons. The reward function for DDPG is designed as
 \begin{displaymath}
 r(s, a) = \begin{cases}
  -100, \ \ \text{$if \ s \in X_u$}\\
 20 - 0.25*|s-150| - 0.75*|v-40|, \ otherwise
\end{cases}
 \end{displaymath}
 The total steps in each epoch is 200. The reward threshold is 1250 for the last 100 epochs, with the condition that the final state is in the goal region. 
 
 \textbf{Learned controllers:} With our approach based on geometric distances, the learned controller is 
 \begin{displaymath}
 \kappa(s, v) = 0.659s - 2.377v
 \end{displaymath}
 
With our approach based on Wasserstein distances, the learned controller is
 \begin{displaymath}
 \kappa(s, v) = 0.868 s -3.035v
 \end{displaymath}
 
 The baseline LQR controller is 
 \begin{displaymath}
 \kappa(s, v) = 0.3162 s - 0.6789 v - 12.3
 \end{displaymath}

\begin{figure}[htbp]
\centering
    \includegraphics[width=0.6\linewidth]{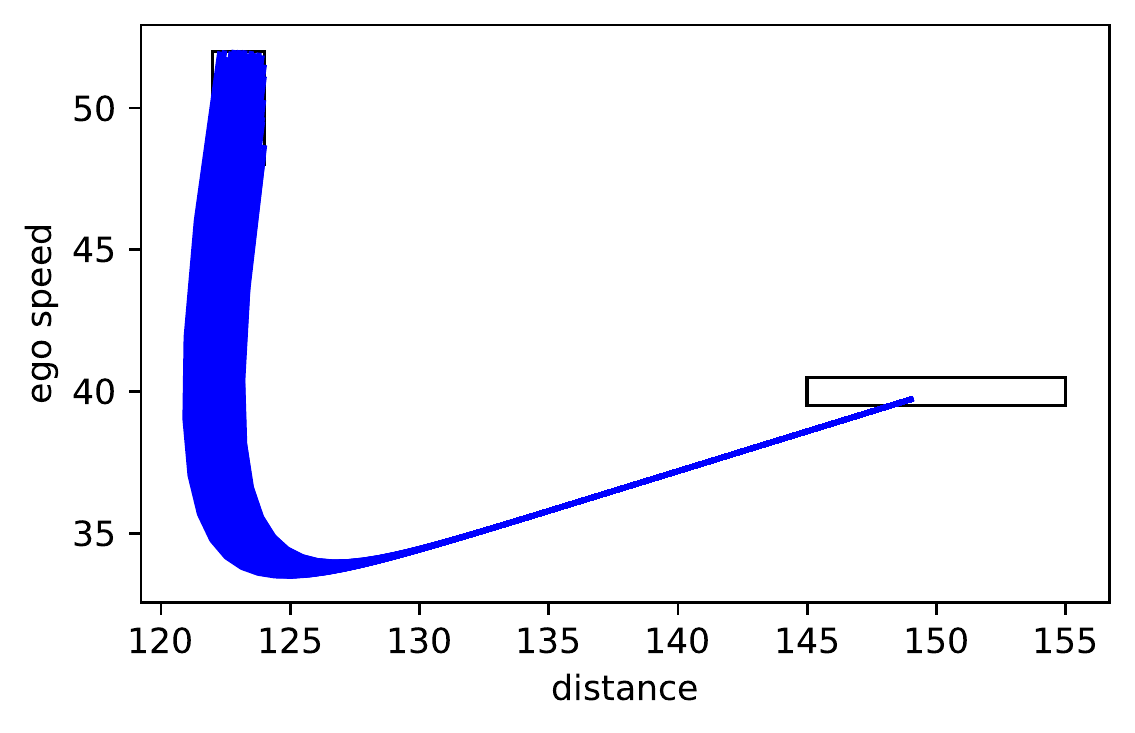}
    \caption{The simulation traces with learned controller $0.789s - 3.085 v$ for ACC in the Webots environment. We can see that the system under this controller can satisfy the reach-avoid property. }
    \label{fig:acc_webots_traces}
\end{figure}

The simulation traces with one of the learned linear controller in Webots is shown in Fig.~\ref{fig:acc_webots_traces}. We can see that the reach-avoid property is satisfied.

\begin{figure}[htbp]
\centering
     \includegraphics[width=0.6\linewidth]{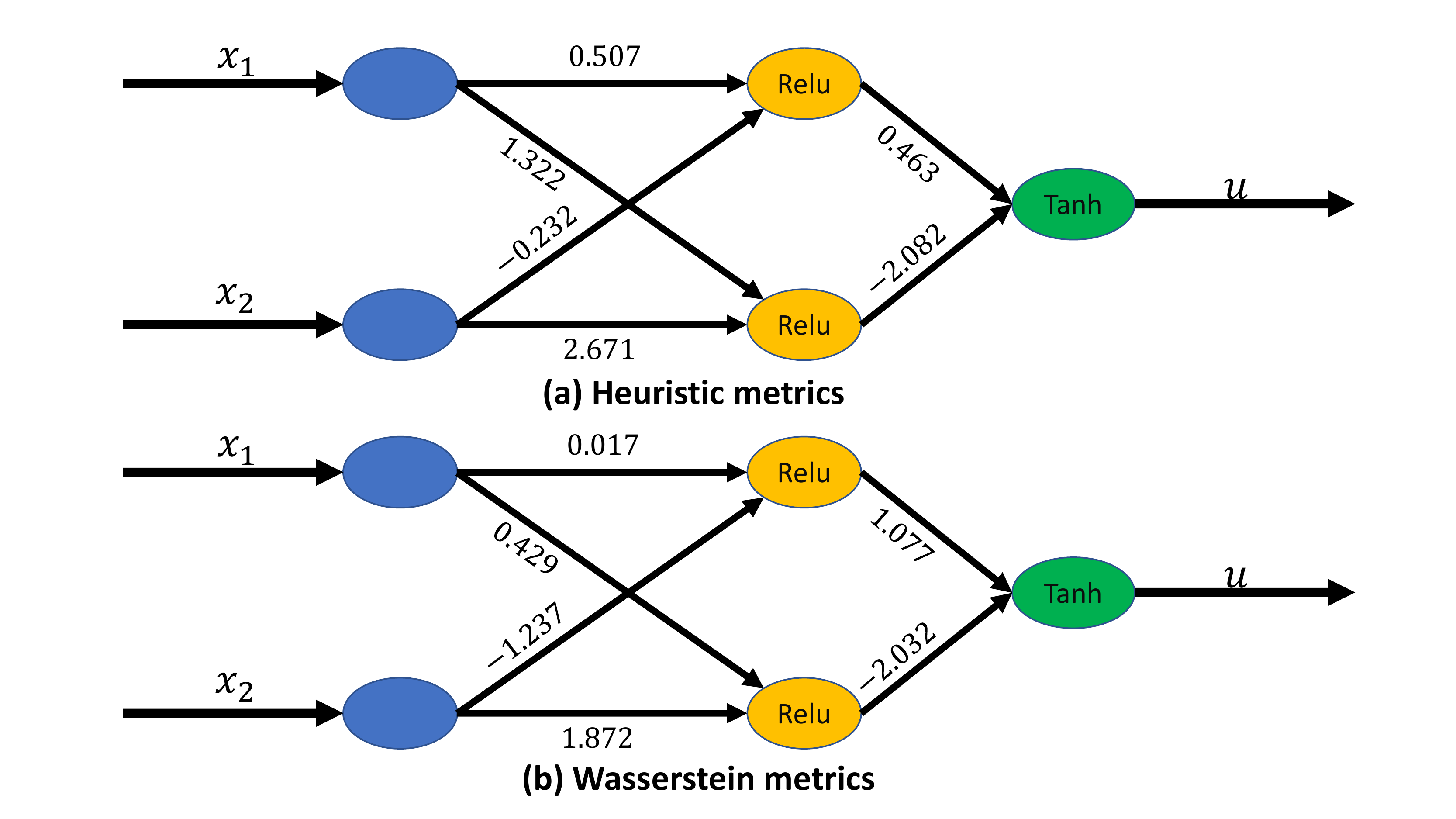}
     \caption{Learned neural network controllers with parameters searched based on our approach with geometric distances (a) and with Wasserstein distances (b) for the oscillator system.}
     \label{fig:OS_nn_plot}
\end{figure}

\textbf{Oscillator:} We propose to learn a tiny neural network controller for this system, with 1 hidden layer of 2 neurons. The parameters of the network are initialized with uniform distribution $[0, 1]$. 
The time horizon of the reachable set computation is set to 15 steps.
The gradient step length is set to 1 for $d_\theta^{g}$, 5 for $d_\theta^{u}$ with geometric distances, and 1 for both Wasserstein distances $W(r, g)$ and $W(r, u)$. The perturbations $p$ on controller parameter are randomly and uniformly sampled 
from $[-0.2, 0.2]$ for safety and $[-0.1, 0.1]$ for goal-reaching of the neural network controller. Following similar ideas of ACC, the baseline DDPG reward function for oscillator is designed as 

 \begin{displaymath}
 r(x_1, x_2) = \begin{cases}
  -50, \ \ \text{$if \ (x_1, x_2) \in X_u$}\\
 5 - 10*|x_1| - 10*|x_2|, \ otherwise
\end{cases}
 \end{displaymath}
The total steps in each epoch is set to 200 steps. The reward threshold is 750 for the last 100 epochs. 

\textbf{Learned controllers for Oscillator}: With our approach, the learned controllers are shown in Fig.~\ref{fig:OS_nn_plot}.
 
 \subsection{Additional Discussions}
 In the Section 5 of the main paper, we discussed hybrid control learning and verification tightness. Here we provide additional details.

\begin{figure}
    \centering
    \includegraphics[width=0.6\linewidth]{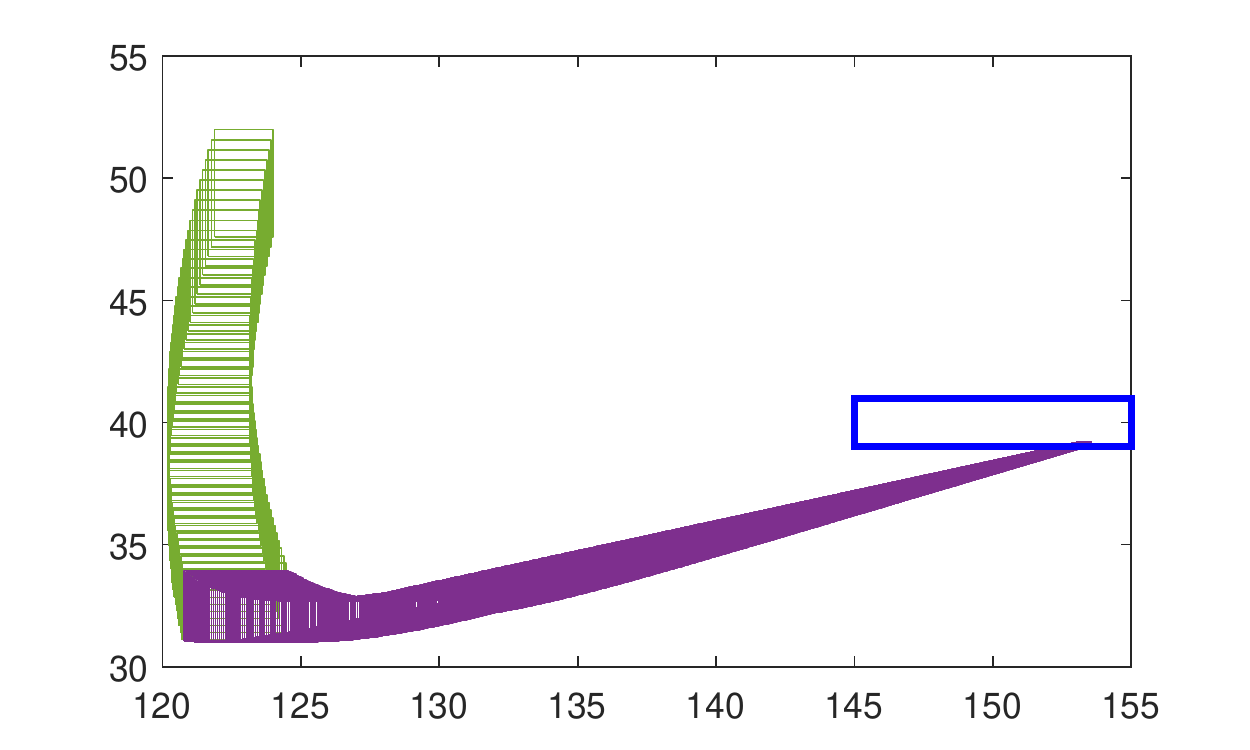}
    \caption{Hybrid control learning with a neural network controller (green) to avoid the unsafe region and a linear controller (purple) to approach the goal region.}
    \label{fig:hybrid}
\end{figure}

\textbf{Hybrid Control Learning:} The failure happens when we want to learn a neural network controller with 75 steps reachable set computation in the ACC system. ACC requires the controller to have a large Lipschitz constant, resulting in too large over-approximation errors. To mitigate this,
we learn a neural network controller for safety and a linear controller for goal-reaching in ACC, as a hybrid control mechanism to mitigate the failure of our approach when facing too-large over-approximation of the reachable set (shown in Fig.~\ref{fig:hybrid}). Specifically, in this example, we compute 6 steps of the reachable set for the neural network controller with the approximated gradients of the distance between the unsafe set. We then treat the last reachable set of the learned neural network controller as the initial set of the linear controller, and compute 50 steps of the reachable set for goal-reaching learning. 

\textbf{Verification Tightness:} To compute tighter verification (i.e., tighter reachable set computation) in ReachNN, we set the CutoffThreshold as $1e-10$, QueueSize as $1000$, and degree bound of Bernstein polynomials as $[2, 2]$. For less tight verification, we set $1e-7$, $300$, and $[1, 1]$ for CutoffThreshold, QueueSize, and degree bound, respectively. For Wasserstein distance on the oscillator system, the tighter reachable set computation in average takes around 40 steps with about 115 seconds for each step to learn a neural network controller, compared to the less tight computation that takes about 55 iterations with 86 seconds for each step. We can see that the overall impact on verification efficiency is not trivial, since while tighter verification may increase the runtime for each step, it may reduce the total number of required steps. We plan to investigate this further in the future work.
\end{document}